\documentstyle[epsf,psfig,aps,prl,twocolumn]{revtex}
\begin{document}
\vskip 0.5cm
\title{
A redefinition of concurrence and its generalisation to bosonic subsystems of
$N$ qubit systems
}
\author{D. D. Bhaktavatsala Rao and V. Ravishankar}
\vskip 0.5cm
\address{Department of Physics, IIT Kanpur, Kanpur-208 016, INDIA
}
\maketitle
\begin{abstract}
We refine the notion of concurrence in this paper by a redefinition of the concept.
 The new definition is simpler, computationally
straight forrward, and allows the concurrence to be directly read off from the state.
It has all the positive
features of the definition given by Wootters over and above which it can discriminate
between different systems to which the Wootters prescription would assign the same value.
 Finally, the definition leads to  a natural extension
of the notion to multiqubit concurrence, which we illustrate with
examples from quantum error correction codes.
\end{abstract}
\maketitle
\narrowtext
\section{Introduction}
Quantum entanglement (QE) is the central feature of quantum mechanics that distinguishes
a quantum system from its classical counterpart. It is also the
corner stone on which many of the novel applications of quantum mechanics -to
 quantum computation, quantum information theory, quantum cryptography and
quantum teleportation - are based. Indeed, it is this promise
that has led to a renewal of interest in QE in recent years.

The simplest example of QE is afforded by a bipartite system of two spin half
(or pseudo spin half) particles, where the Hilbert space has the minimal dimension four.
This system is also called a two qubit system (2QS) in the context of
 quantum computations.
 For a 2QS,
there is essentially only one measure of entanglement; it may be given
{\it e.g.}, by the von-Neumann entropy ${\cal E}_N$, or expressed in terms of
 the sum of bilinears in the eigen values as given by the quantity
 $1 -Tr \rho^{(r)2} \equiv {\cal E}_{tr}$, or in terms of det$\rho^{(r)} = {\cal E}_d$.
  $\rho^{(r)}$ is the reduced density
 matrix obtained by taking a marginal trace over one of the spin degrees of freedom.
 States with a vanishing entanglement are separable (in the strong sense):
 they admit a factorisation
 $\vert 1,2> = \vert 1> \otimes \vert 2>$. On the other hand, fully entangled states
 have the maximum correlation, and are collectively designated as Bell states.

  Consider now a N-qubit system (NQS),
 realised as  a multipartite system of $N$ spin halfs. NQS upto $N=4$ have
 been prepared
experimentally  with photons, for example \cite{exp1}.
 It is easy to check that for  an NQS,
${\cal E}_N$, ${\cal E}_{tr}$ and ${\cal E}_d$ are not equivalent.  Further, none of them
is exhaustive. A natural question that arises is how one may identify a complete set
of entanglement measures for any NQS. There have been several proposals \cite{meyeis}
that attempt at answering this question. One note worthy proposal is from Sixia Yu {\it et al}
\cite{sixia}
who posit a hierachy of $N-1$ classes of entanglement. The total number grows exponentially
with increasing $N$.

An alternative approach is to look for   physically interesting measures,
apart from  ${\cal E}_N$ and ${\cal E}_{tr}$,
with
(potential) applicability  to quantum computation.
This simpler approach would also give direction to the experimentalists to prepare
NQS in specified states.

In this context, a useful question to ask is what the entanglement property
of a 2QS is, when looked upon as a subsystem of an NQS, when $N \ge 3$ (There
would be no new information when $N=2$).  There would
be $NC_2$  such subentanglements, as analogues of two particle correlations. In defining
the new measure, we
seek to determine the  'proximity' of a given 2QS  to the classic
Bell states mentioned above. 

Wootters \cite{woot} has developed a closely related concept called concurrence
(hereafter denoted ${\cal C}_W$) for 2QS,
which is required to characterise the so called 'entanglement of formation' (EOF) \cite{hw}. 
 Indeed, EOF
 is defined to be the minimum entropy carried by a mixed state, under all possible
resolutions 
\begin{eqnarray}
\rho =\sum_i p_i \vert \psi_i><\psi_i \vert
 \end{eqnarray}
 of a reduced density matrix
of a 2QS; 
the set 
\{$\vert \psi_i><\psi_i \vert$\} does not {\em necessarily}  form a basis - much less
an orthonormal set. ${\cal C}_W$, which measures this EOF is then defined as follows:
\newtheorem{definition}{Definition}
 \begin{definition}[Wootters:]
 The concurrence for a mixed 2QS is given by
 \begin{eqnarray}
 {\cal C}_W = max \{0, \lambda_1-\lambda_2-\lambda_3-\lambda_4\} \nonumber
 \end{eqnarray}
 where the $\lambda_i$ are the eigenvalues, arranged in decreasing order of magnitude,
 of the operator \\
 $R=(\sqrt{\rho}\tilde{\rho}\sqrt{\rho})^{1/2};\,\, \tilde{\rho} = \sigma_y \otimes \sigma_y
 \rho \sigma_y \otimes \sigma_y.$
 \end{definition}
 The so called EOF is itself given by
 \newtheorem{proposition}{Proposition}
 \begin{proposition}
 The entanglement of formation is given by \\
 ${\cal{E}}(C_W) = h \left ( \frac{1 + \sqrt{1 - C_W^2}} {2} \right );
h(x) = -x{log_2}x - (1-x){log_2}(1-x).$
\end{proposition}
The above definitions are attractive, and have been used to estimate bipartite entanglement
in spin systems \cite{sub}. There are also reports that states with nonvanishing ${\cal C}_W$
have been prepared experimentally \cite{expt2}. 
Attempts have been made to generalise the definition
to higher spin sectors \cite{gen}. There is, thus, quite an interest in ${\cal C}_W$.

The definition of ${\cal C}_W$, rather involved in its form, crucially hinges on the
concept of EOF. However, it is well known that
there is no way of inferring how a system is prepared \cite{terhaar}, no matter
which property of the density matrix we look at, unless it  corresponds to a pure
state. Thus the physical interpretation
of EOF is rather dubious, and does not seem to have any
operational significance.
Indeed, suppose the expansion of $\rho$ in Eq.1 is replaced by a more general decomposition
$\rho =
\int_{\{\alpha_i\}}[d \alpha_i] p_{\{\alpha_i\}}
\vert \psi(\{\alpha_i \})><\psi(\{\alpha_i\}) \vert$,
where the states are now labelled by a set of continuous parameters $\{\alpha_i\}$ with
$[d \alpha_i]$ being an appropriate integration measure. The parameters could be taken to be
the coordinates
of the associated compact phase space, for instance.
The probabilities '$p(\{\alpha_i\})$' have as much physical significance
 as the discrete variables $p_i$ in Eq.1. But the new expansion
 would change the criterion for EOF, leading to a different
expression for concurrence.

With this critique of ${\cal C}_W$ in mind, we propose an alternative definition of
concurrence ( hereafter called ${\cal C}_R$), with a two fold aim: to attain conceptual
and computational simplicity. ${\cal C}_R$
 is required merely to capture the entanglement of
a 2QS which is a subsystem of a larger NQS, in a manner that suggests a
generalisation to higher spin sectors. We take this up in the next section.

\section{Concurrence}
 As a warm up, and for motivation, consider a 2QS first.
  Let a 2QS be in the state
\begin{eqnarray}
  \vert1,2> = \alpha \vert \uparrow \uparrow  > +
  \alpha_2 \vert \uparrow \downarrow >
              + \alpha_3 \vert \downarrow \uparrow > +
	       \alpha_4 \vert \downarrow \downarrow >.
\end{eqnarray}
  The key step in defining concurrence is in recognising that
 ${\cal E}_c = 2 \vert \alpha_1 \alpha_4 - \alpha_2 \alpha_3 \vert$ 
 is a measure of entanglement,
 equivalent to the standard measures listed above.
  ${\cal E}_c$ is much simpler to evaluate, though.
 Our  notion of concurrence is inspired by,
 and is very close in its definition
 to, ${\cal E}_c$. Although ${\cal E}_c$ does {\it not}
 carry any new information,
 it does exhibit a symplectic structure
 which is not apparent in the other two measures. Further, as observed by Hill and Wootters
 \cite{hw}, it can also be written
 in the form
$ \vert <\{12\}  \vert 1,2 > \vert$,
the inner product of $\vert 1,2>$ with its conjugate (equivalently, time reversed)
state which we define thus:
\begin{eqnarray}
\vert \{12\}> = \sum_{m_1,m_2} (-1)^{m_1+m_2} \alpha_{m_1,m_2}^{\star}
 \vert -m_1, -m_2>.
\end{eqnarray}
The main thrust of the paper is that this form of entanglement for a 2QS
 needs {\it little} modification in defining concurrence in NQS, $N \ge 3$.
 The
definition will be given in steps, so as not to have a cluttered notation.

\noindent {\bf The 3QS:} Consider the simplest case, of a 3QS, prepared in a pure state.
 Let
 \begin{eqnarray}
 \vert 1,2,3 > = \sum_{m_1,m_2,m_3} \alpha_{m_1,m_2,m_3} \vert m_1,m_2,m_3>.
 \end{eqnarray}
 We define:
  \begin{definition}
  The concurrence between the first two qubits of a 3QS
 is given by
 \begin{eqnarray}
  {\cal C}^{(12)}_R = \vert <\{12\},3 \vert 1,2,3> \vert;
  \end{eqnarray}
 % where
  $\vert \{12\}3 >
 \equiv \sum \ (-1)^{m_1+m_2} \alpha_{m_1 m_2 m_3}^{\star} \vert -m_1, -m_2, m_3 >.$
 \end{definition}
 which is obtained by a partial conjugation -- on the sector (12).
 The definition of concurrences ${\cal C}^{(13)}$ and ${\cal C}^{(23)}$ follow similarly.
 By definition ${\cal C}^{(ij)}_R$ take values in the range $[0,1]$.
${\cal C}^{(12)}_R$ also has an in built
symplectic structure; to see this, consider the quadruplet of vectors
\begin{eqnarray}
\{{\cal V}_1 =(\alpha_1, \alpha_2), \;
{\cal V}_2 =(\alpha_3, \alpha_4),  \nonumber \\ {\cal V}_3 =(\alpha_5, \alpha_6), \;
 {\cal V}_4 =(\alpha_7, \alpha_8) \}; \nonumber
 \end{eqnarray}
  the expansion is in the ordered basis $\{ \vert \uparrow \uparrow \uparrow>,
  \vert \uparrow \uparrow \downarrow>, \vert \uparrow \downarrow \uparrow>,
  \vert \uparrow \downarrow \downarrow>,
  \vert \downarrow \uparrow \uparrow>, \vert \downarrow \uparrow \downarrow>,
  \vert \downarrow \downarrow \uparrow>, \vert \downarrow \downarrow \downarrow> \}$.
  The concurrence is now simply written as
 \begin{eqnarray}
 {\cal C}^{(12)}_R = 2\vert \tilde{\cal V}_4{\cal V}_1 - \tilde{\cal V}_3{\cal V}_2 \vert,
 \end{eqnarray}
 which
 naturally generalises the definition of entanglement in a 2QS in a manner appropriate to
 our purpose \cite{fn}. $\tilde{{\cal V}}$ is the transpose of ${\cal V}$.
 One can straight away construct quadruplets for ${\cal C}^{(23)}_R, {\cal C}^{(13)}_R$
 by inspection. It is also straight forward to generalise
 the definition when the parent state is an NQS in a pure state. We write the expression
 for the $(12)$ sector:

\begin{definition}
 The concurrence of the first two qubits of a NQS in a pure state
 $\vert 1,2,\cdots,N> = \sum \alpha_{m_1,m_2, \cdots,m_N}\vert m_1,m_2,\cdots,m_N >$
 is given by $\vert < \{12\}, 3, \cdots , N \vert \vert 1,2,\cdots,N>\vert$, where the
 conjugate state is now defined to be
 $\vert \{12\}, 3, \cdots , N> = \sum_{i=1}^N \alpha{\star}_{-m_1,-m_2, m_3, \cdots , m_n}
  \vert 1,2,\cdots,N>$.
  \end{definition}

 The computational simplicity of ${\cal C}_R$ is evident, from its very form, if only
 we demonstrate its viability. A convenient method is to use ${\cal C}_W$ as a benchmark
 to test the new definition against.
 For, the definition of ${\cal C}_W$ is also motivated by the
 entanglement measure ${\cal E}_c$.
 Figure 1. gives a comparison of the
 two definitions,
 where the states are generated randomly. One sees an overall agreement
 so long as ${\cal C}_W$
 remains positve, without having to impose the minimality condition by hand. In those
 regions where ${\cal C}_W$ has to be declared as vanishing (when $\lambda_1-\lambda_2-
 \lambda_3-\lambda_4 < 0 $) in  {\it Definition 1}, ${\cal C}_R$ differs from ${\cal C}_W$:
 Indeed,
 these states are distinguishable, in the sense that each of them has a preferential proximity
 to one of the Bell states which is more clear in Fig.2.

  The computational advantage is of course evident
 from the form of ${\cal C}_R$. For, If we start with a NQS,
 there are $NC_2$ such concurrences.
 Even in regions where ${\cal C}_W$ has no interpretational problem, {\it Definition 1}
 requires an evaluation of three reduced density matrices, a determination of their
 partner states, and an evaluation of the eigenvalues of $R$. The necessity of these
  computations
 gets obviated in the
 determination of ${\cal C}_R$.
\vspace{0.6cm}
\begin{figure}[hpb]
\vspace*{-0.6cm}
\centerline{\psfig{figure=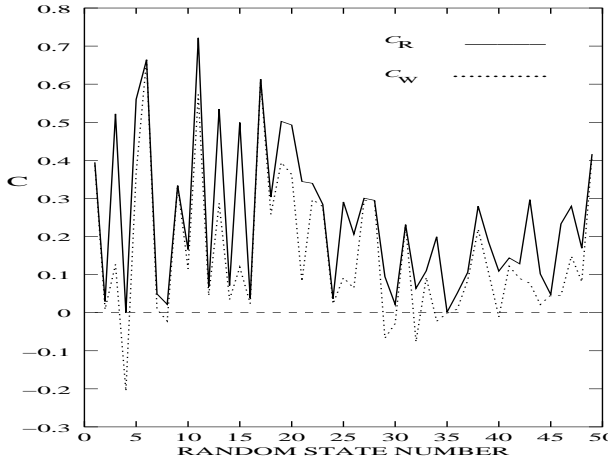,width=8cm,height=6cm}}
\vspace{0.2cm}
\caption{${\cal C}_R$ compared with ${\cal C}_W$ in the two particle
sector for randomly generated
4 particle states.
They show a relative monotonic behaviour  except where
${\cal C}_W$ vanishes. The negative values get reset to zero in ${\cal C}_W$.}
\end{figure}
\vspace{0.4cm}

\vspace{0.3cm}
\begin{figure}[hpb]
\vspace*{-0.6cm}
\centerline{\psfig{figure=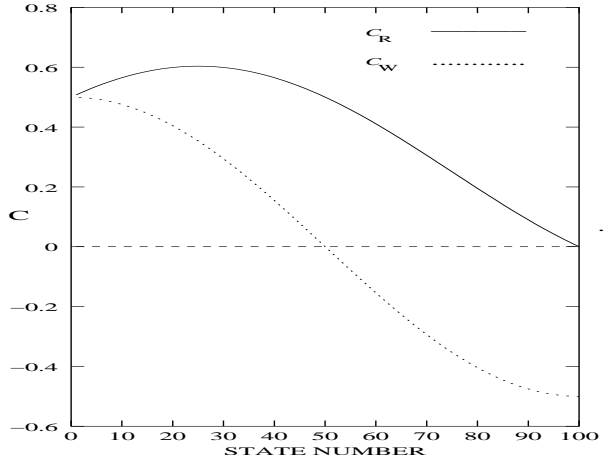,width=8cm,height=6cm}}
\vspace{0.2cm}
\caption{${\cal C}_R$ contrasted with ${\cal C}_W$ in the two particle sector
for 4 particle states with
continously varying parameters. The negative values are reset to zero in ${\cal C}_W$.}
 
\end{figure}
\vspace{0.4cm}

We proceed to enlarge the definition of concurrence to higher spin sectors.
Note that there is no analogue of ${\cal C}_W$ here,
since the criterion for EOF which was evolved
in \cite{woot} is specific to a 2QS.
The generalisation in our case is naturally
 suggested by the form for ${\cal C}_R$ written above. Indeed, let
 \begin{eqnarray}
 \vert 1,2,\cdots,N> = \sum \alpha_{m_1,m_2, \cdots,m_N}\vert m_1,m_2,\cdots,m_N >
 \end{eqnarray}
 be the state in which an NQS is. We define the conjugate state
 \begin{eqnarray}
 \nonumber
 &&
 \vert \{12\cdots k\} k+1,\cdots,N> =
 \sum (-1)^{\sum_i^km_i} \alpha_{m_1,m_2, \cdots,m_N}^{\star} \\
 &&
 ~~~~~~~~~~~~~~~\vert -m_1,-m_2,\cdots,-m_k, m_{k+1},\cdots m_N >
 \end{eqnarray}
  \begin{definition}
  The concurrence in the first $k$ qubits of an NQS is given by
  \begin{eqnarray}
  {\cal C}^{(12 \cdots k)}_R = \vert <\{12\cdots k\}k+1,\cdots,N \vert 1,2,\cdots ,N> \vert.
  \end{eqnarray}
  \end{definition}

  The above definition is unfortunately restrictive;
  It works only when the subsystem of the NQS is  bosonic;
  The fermionic states ($k$ odd) are always orthogonal to their conjugates.
  Apart from this,
  there is  generality with respect to the choice of the $k$ qubits since any
  $k$ qubits may be chosen to be in the order given, by a permutation.
    The number of independent concurrence measures
  is given by $NC_k$.
   \vspace{0.4cm}
\begin{figure}[htp]
\vspace*{-0.6cm}
\centerline{\hbox{\psfig{figure=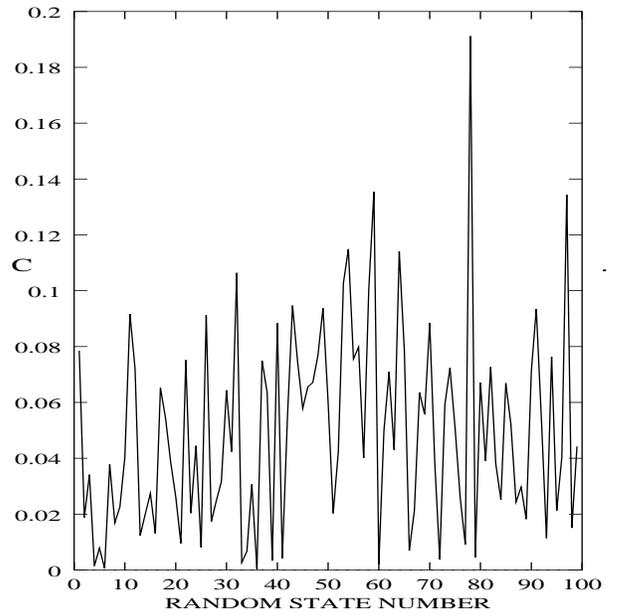,width=8cm,height=8cm}}}
\vspace{0.2cm}
\caption{4Q concurrence for randomly generated 6 particle states with the new definition.}
\end{figure}
\vspace{0.4cm}

  Figure 3 displays the 4Q concurrences for states generated from
  a 6QS.  As an illustration, and as an example of the discerning capability of the higher order
  concurrences which we have defined here, the Shor code will be compared with the Steane
  code, both of which are used in quantum error correction \cite{NC}.
  The Shor code  $\vert 0_L> (\vert 1_L>)$ is a nine qubit
  state, written as a direct product of three 3Q states each of which have the form
  $\vert 000> \pm \vert 111 >$, in writing which we have employed the qubit notation:
  $0 \rightarrow \downarrow  (-\frac{1}{2}), 1 \rightarrow \uparrow
  (\frac{1}{2})$.
  This entangled
  state has vanishing concurrences in {\em all} orders. In contrast, the Steane code,
  constructed
  with seven qubits has a more involved structure; it is not seperable the way the Shor state
  is., and is given by
  $\vert 0_L> =\frac{1}{\sqrt{8}}(\vert 0000000> + \vert 1010101> +\vert 1100110> +
    \vert 0001111> + \vert 0110011>
   + \vert 1011010> + \vert 0111100> + \vert 1101001>)$.
   As a manifestation, although all the 2Q concurrences vanish, the  4Q concurrences
  survive in the  sectors $\{ (1247), (1256), (1346), (1357), (2345), (2367), (4567) \}$, and
  all of them attain the maximum allowed value 1.
  The 6Q concurrences  vanish. These  features highlight quantitatively the manner in which
  the two codes differ from each other.   At this point, it is good to remember that a complete separability in the strong sense,
  where the multiqubit state is a direct product of single qubit states, necessarily
  implies a vanishing concurrence of all orders. The converse is not necessarily true, as exemplified by the state generating the Shor code.

  Finally, the question of handling mixed states still remain. Not getting into the general
  case, only the 2Q concurrence will be discussed. The general case only involves rewriting
  the argument with more indices. Indeed, given a 2QS in a mixed state $\rho$,
  the method is to look upon $\rho$ as having descended from a pure parent state of a higher
  dimension $N$. Significantly, $N \le 4$. The determination of parent state is done by simple
  inspection. It also follows that if $\rho$ has only real components, then
  ${\cal C}_R = 2{\vert} {\rho}_{14} -{\rho}_{23} {\vert}.$
  It is not difficult to check that although the parent
  state is not unique, the concurrence determined will be so - reflecting the fact that
  it is a property of the system, and not of the parent - introduced purely
  as an intermediate step. This final remark completes the programme undertaken.

  In conclusion, we have redefined in this note the concept of concurrence which has
  a much simpler form, and is computationally trivial - compared to ${\cal C}_W$ and its
  generalisations thereof. It is not based on
  the notion of EOF. While concurrences do not completely characterise a system - which was
  not our aim - it appears that they do constitute an important subset of entangled states,
  especially in the two qubit sector. The higher order concurrences also have a significance, as
  illustrated by the concurrence properties of Shor and the Steane codes.
  The true import  of entanglements of this kind would manifest
  with only a more geometric approach, which will be taken up in a subsequent publication.
  
  It is our great pleasure to thank V. Subrahmanyam for many discussions on this
  topic.


\begin{references}
\bibitem{fn} Clearly, elements of $\rho^{(r)}$ in the (12) sector
 are given by $\rho^{(r)}_{ij} = {\cal V}_j^{\dagger}{\cal V}_i$ which, incidentally gives
 a three parameter family of three qubit states all of which reduce to the same density
 matrix in the (12) sector. The $SU(2)$ structure is obvious.
\end{references}
\end{document}